\documentclass[manuscript]{acmart}

\usepackage{tabularx}
\usepackage{enumitem}
\usepackage{multirow}
\usepackage{subcaption}
\usepackage{soul}

\usepackage[normalem]{ulem}

\AtBeginDocument{%
  \providecommand\BibTeX{{%
    \normalfont B\kern-0.5em{\scshape i\kern-0.25em b}\kern-0.8em\TeX}}}

\setcopyright{none}
\copyrightyear{2024}
\acmYear{2024}
\acmDOI{XXXXXXX.XXXXXXX}

\acmConference[CHI '24]{Make sure to enter the correct
  conference title from your rights confirmation email}{May 11--16,
  2024}{Honolulu, HI, USA}

\acmBooktitle{CHI '24: ACM CHI conference on Human Factors in Computing Systems,
 May 11--16, 2024, Honolulu, HI, USA} 
\acmPrice{15.00}
\acmISBN{978-1-4503-XXXX-X/18/06}

\begin{document}

\title{Design Principles for Generative AI Applications}

\author{Justin D. Weisz}
\email{jweisz@us.ibm.com}
\orcid{0000-0003-2228-2398}
\affiliation{
    \institution{IBM Research AI}
    \city{Yorktown Heights}
    \state{NY}
    \country{USA}
}

\author{Jessica He}
\email{jessicahe@ibm.com}
\orcid{0000-0003-2368-0099}
\affiliation{
    \institution{IBM Research AI}
    \city{Seattle}
    \state{WA}
    \country{USA}
}

\author{Michael Muller}
\email{michael_muller@us.ibm.com}
\orcid{0000-0001-7860-163X}
\affiliation{
    \institution{IBM Research AI}
    \city{Cambridge}
    \state{MA}
    \country{USA}
}

\author{Gabriela Hoefer}
\email{ghoefer@ibm.com}
\orcid{0000-0002-9881-1244}
\affiliation{
    \institution{IBM}
    \city{New York}
    \state{NY}
    \country{USA}
}

\author{Rachel Miles}
\email{Rachel.Miles@ibm.com}
\orcid{0009-0001-7493-5111}
\affiliation{
    \institution{IBM}
    \city{San Jose}
    \state{CA}
    \country{USA}
}

\author{Werner Geyer}
\email{Werner.Geyer@us.ibm.com}
\orcid{0000-0003-4699-5026}
\affiliation{
    \institution{IBM Research AI}
    \city{Cambridge}
    \state{MA}
    \country{USA}
}

\renewcommand{\shortauthors}{Weisz et al.}

\begin{abstract}
   Generative AI applications present unique design challenges. As generative AI technologies are increasingly being incorporated into mainstream applications, there is an urgent need for guidance on how to design user experiences that foster effective and safe use. We present six principles for the design of generative AI applications that address unique characteristics of generative AI UX and offer new interpretations and extensions of known issues in the design of AI applications. Each principle is coupled with a set of design strategies for implementing that principle via UX capabilities or through the design process. The principles and strategies were developed through an iterative process involving literature review, feedback from design practitioners, validation against real-world generative AI applications, and incorporation into the design process of two generative AI applications. We anticipate the principles to usefully inform the design of generative AI applications by driving actionable design recommendations.
\end{abstract}



\begin{teaserfigure}
  \centering
  \includegraphics[width=0.90\textwidth]{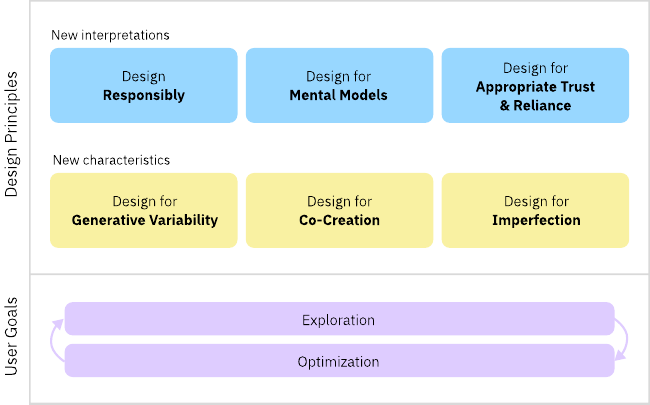}
  \caption{Six principles for the design of generative AI applications. Three principles offer new interpretations of known issues with AI systems through the lens of generative AI, and three principles identify unique characteristics of generative AI systems. The principles support two user goals: optimizing a generated artifact to satisfy task-specific criteria, and exploring different possibilities within a domain.}
  \Description{Six principles for the design of generative AI applications. Three principles offer new interpretations of known issues with AI systems through the lens of generative AI, and three principles identify unique characteristics of generative AI systems. The principles support two user goals: optimizing a generated artifact to satisfy task-specific criteria, and exploring different possibilities within a domain.}
  \label{fig:principles-schematic}
\end{teaserfigure}


\maketitle

\section{Introduction}

Generative AI technologies have reached an inflection point in consumer adoption and enterprise value, sparked by technological advancements in machine learning architectures such as GANs~\cite{goodfellow2014generative, karras2019style}, VAEs~\cite{kingma2013auto}, and transformers~\cite{vaswani2017attention, devlin2018bert}. Models such as StyleGAN~\cite{karras2019style}, GPT~\cite{radford2018improving, radford2019language, brown2020language, openai2023gpt4}, and Codex~\cite{chen2021evaluating} have demonstrated that powerful generative models can produce works at a human-like level of fidelity. Today, consumer applications such as ChatGPT\footnote{ChatGPT. https://chat.openai.com}, DreamStudio\footnote{DreamStudio. https://beta.dreamstudio.ai}, and DALL-E\footnote{DALL-E. https://labs.openai.com} are making these technologies widely available and setting the bar for people's expectations of what generative AI can do. Startups such as Cohere\footnote{Cohere. http://cohere.ai} and Anthropic\footnote{Anthropic. http://anthropic.com} are reducing the friction of embedding large language models in consumer applications. Enterprises such as IBM, Microsoft, Amazon, and Google are creating platforms for businesses to infuse generative technologies into their products and services.

This commercialization of generative AI technologies is fueled by the ultra-rapid development of large-scale foundation models~\cite{bommasani2021opportunities} that reduce the time and costs for developing generative AI systems. However, much attention in machine learning research communities has focused on developing advancements to the \emph{technology}: scaling model parameter counts~\cite{singh2023effectiveness, lepikhin2020gshard}, evaluating model performance~\cite{liang2022holistic, zheng2023judging, srivastava2022beyond}, tuning models efficiently to perform new tasks~\cite{wang2023multitask, chen2023parameter}, and aligning models~\cite{ouyang2022training, ziegler2019fine} to reduce their propensity to produce speech that is hateful, abusive, profane, or otherwise toxic~\cite{ji2023beavertails, hartvigsen2022toxigen}. Although these advancements serve to improve the state of the art, they do not recognize an important half of what \citet{ehsan2021expanding} call the ``human-AI assemblage'' -- the \emph{human}.

Generative models have enabled a radically new way for people to interact with computing technologies. People are now able to craft specifications for the kinds of outputs they desire, such as via natural language prompts, and generative models are able to produce outputs that conform to those specifications. \citet{nielsen2023ai} recently identified this form of interaction as \emph{intent-based outcome specification} and argued that it is the first new UI interaction paradigm in 60 years. This form of interaction is fundamentally different from previous interaction paradigms (e.g. punchcards, command line interfaces, and graphical user interfaces), because it shifts control over how computation is performed away from the user and toward generative AI models. With this shift in control, how are we to design user experiences that help people interact with generative AI applications in effective and safe ways?

Over at least the past four decades, researchers and practitioners within human-computer interaction (HCI) have produced numerous guidelines, principles, practices, and frameworks for the design of effective and safe computing systems. Some guidelines are presented as generally applicable to most kinds of interactive computing systems, such as \citeauthor{nielsen1990heuristic}'s heuristics~\cite{nielsen1990heuristic} and \citeauthor{shneiderman2016designing}'s strategies for designing effective human-computer interaction~\cite{shneiderman2016designing}. Other design guidelines are technology-specific, such as \citeauthor{bevan2005guidelines}'s guidelines for web usability~\cite{bevan2005guidelines} and various guidelines for AI systems and human-AI interaction~\cite{amershi2019guidelines, pair2021people, ibm2022aiethics}. However, none of the guidelines developed within the HCI community have yet addressed the specific nuances present with generative AI.

In this paper, we build upon the HCI tradition of distilling cumulative knowledge into a practical set of principles, specifically for the design of generative AI user experiences (UX). We opted to produce a set of \emph{principles}, rather than \emph{guidelines}, to highlight their fundamental nature to the design of generative AI UX. 

Our paper makes the following contributions to the CHI community:

\begin{itemize}
    \item We \textbf{introduce} a set of six principles for the design of generative AI applications (Table~\ref{tab:design_principles}). Three principles identify new considerations for generative AI, and three offer new interpretations of known issues with AI systems. Each principle is coupled with a set of practical strategies and examples (Appendix~\ref{appendix:extended-description}) for how to put the principle into practice -- either through the design process itself or through specific UX capabilities -- in order to aid design practitioners\footnote{We consider a design practitioner to be anyone involved in making deliberate decisions that impact the design of an application, including user researchers, interaction designers, visual designers, product managers, and more.} in applying the principles to real-world UX. 
    
    \item We establish the \textbf{practical value} of the design principles by conducting a rigorous and systematic validation that included multiple rounds of testing, feedback collection, iteration, and use by two generative AI application design teams.
    
    \item We \textbf{provide insight} on how we made decisions regarding the organization of the principles, navigated issues of conceptual redundancy and overlap, and fostered adoption of the principles within our organization.
\end{itemize}

\section{Related Work}
Over the past four decades, the HCI community has conducted numerous examinations of user interface design for computing technologies and distilled best practices into various forms of design guidelines. We identify and review two categories of these guidelines: those that apply to the general UX design of computing systems and those that apply specifically to AI-infused systems.

\subsection{Guidelines for human-computer interaction and user interface design}
\label{sec:guidelines-hci}

Design guidelines have a rich history in HCI research. We provide a brief historical sketch of human-computer interaction design guidelines and their evolution as a result of new paradigms in user interface technologies.

\citeauthor{licklider1962line} provide an early example of guidelines for the design of console interfaces in, ``On-line man-computer communication''~\cite{licklider1962line} (reflecting the gendered language of their time). Although their guidelines are not presented in a modernly-recognizable form\footnote{For example, they provide guidelines on allocating tasks between humans and computers that ``exploit the complementation that exists between human capabilities and present computer capabilities''~\cite[p.114]{licklider1962line}.}, their work does introduce ideas about humans' needs when working with computers. Other early forms of human-computer interaction guidelines include those for the Spacelab Experiment Computer Application Software (ECAS)~\cite{dodson1978development}, and for the command terminals of battlefield automated systems~\cite{sidorsky1980guidelines, sidorsky1984design}. Even in these early days of user interface design, it was recognized that, ``[l]acking consistent design principles, current practice results in a fragmented and unsystematic approach to system design, especially where the user/operator-system interaction is concerned.''~\cite[p. v]{sidorsky1984design}.

A significant inflection point in the design of user interfaces for computing technologies came with the rise of personal computing and the concurrent emergence of graphical user interfaces (GUIs) as a new interaction paradigm. New guidelines were needed for the new visual metaphors of windows, icons, menus, and pointers. Apple's Human Interface Guidelines~\cite{apple1987apple} provides a prominent example of practical guidelines for GUI design. \citet{smith1986guidelines} developed more formal guidelines for GUIs in which they described six functional areas: data entry, data display, sequence control, user guidance, data transmission, and data protection.

As \citeauthor{grudin1990computer} described in an influential retrospective analysis \cite{grudin1990computer}, the ``site'' of human-computer interaction began to move away from a terminal in a lab and ``reached out'' into other contexts such as home and office environments. New methods were required to understand and design for these changing circumstances. With heuristic evaluation \cite{nielsen1990heuristic}, Nielsen provided a set of methods for ``discount usability engineering''~\cite{nielsen1995scenarios} that helped designers more easily assess their interfaces, while \citeauthor{lewis1997cognitive}'s cognitive walkthrough method~\cite{lewis1997cognitive} provided a way for designers to conduct more detailed and tailored analysis.

The next technological inflection point that necessitated a shift in design guidelines occurred with the rise of the Web. Unlike the previous decades, web design involved diverse and competing hardware and software. These complexities led to what \citeauthor{mariage2011state} describe as a ``jungle of guidelines [intended to] address many different issues''~\cite[Introduction]{mariage2011state}. One result was that, out of 11 generic web design guidelines, \citeauthor{cappel2007usability} found that a mean of only 5.5 guidelines were followed across 500 companies' websites \cite{cappel2007usability}.
Adding to the diversity, technologically-literate advocates emerged for people with disabilities \cite{leporini2008applying, romen2012validating}, older users \cite{becker2004study, kurniawan2005derived}, and users from diverse cultures \cite{alexander2017cross}. \citet{bevan2005guidelines} acknowledged that the available wealth of guidelines addressed different issues in different ways for different constituencies, and that this situation was likely to continue.

The advent and widespread commercial adoption of smartphones, mobile apps, app stores, and mobile web sites necessitated yet another new design language, optimized for smaller screens and touch-based interactions. Some work in this space focused on developing design frameworks and guidelines for mobile apps and workflows (e.g., \cite{lupanda2021design, pinandito2017analysis, shitkova2015towards, karlson2010mobile, nagata2003multitasking, hoehle2016leveraging}). Guidelines also emerged covering mobile design for more specialized populations, including older users~\cite{alsswey2018towards, chirayus2020cognitive}, users of courseware~\cite{jia2018design}, users from diverse cultures~\cite{alsswey2018towards}, users with disabilities~\cite{park2014toward}, and users with diverse literacies~\cite{srivastava2021actionable}. Guidelines were also developed for specific mobile app domains such as health care~\cite{alnanih2016mapping, jones2016ethical} and finance~\cite{mohan2015mobile, ali2019study, huebner2018people}, as well as ethical concerns around privacy~\cite{li2022understanding_a, li2022understanding_b} and the use of mobile apps for research purposes~\cite{mcmillan2013categorised, rooksby2016implementing}.

\subsection{Guidelines for human-AI interaction}
\label{sec:guidelines-hai}

Within the past few years, the emergence of AI as a design material~\cite{yildirim2022experienced, holmquist2017intelligence, feng2023ux, dove2017ux} has necessitated guidelines that inform its use. A growing body of work within the human-centered AI research community has proposed best practices for human-AI interaction in the form of design guidelines (e.g.,~\cite{amershi2019guidelines, balasubramaniam2020ethical, wickramasinghe2020trustworthy, yildirim2023investigating, liu2022design}), formal studies (e.g.,~\cite{liu2023beyond, calderwood2020novelists}), toolkits (e.g.,~\cite{madaio2020co}), and reviews (e.g.,~\cite{jobin2019global, wright2020comparative, hagendorff2020ethics, weidinger2021ethical}). 

Some of these guidelines include claims of universal applicability\footnote{Multiple scholars have critiqued the concept of ``universal'' design as privileging certain assumed ``normal'' populations~\cite{bardzell2010feminist, trewin2019considerations}.} or being of a general nature to AI-infused systems (e.g.,~\cite{shneiderman2020human, amershi2019guidelines}). Other guidelines focus on specific types of AI technologies (e.g. text-to-image models~\cite{liu2022design}), specific domains of use (e.g. creative writing~\cite{calderwood2020novelists}), or specific issues regarding the use of AI, including ethics~\cite{balasubramaniam2020ethical, hagendorff2020ethics, jakesch2022different, jobin2019global}, fairness~\cite{madaio2020co}, human rights~\cite{fukuda2021emerging}, explainability~\cite{mohseni2021multidisciplinary}, and user trust~\cite{wickramasinghe2020trustworthy}. Finally, as more consumer products incorporate AI technologies, industry leaders including Google~\cite{pair2021people}, Microsoft~\cite{amershi2019guidelines, li2022assessing} and Apple~\cite{apple2022higml} have developed and published their own guidelines; \citet{wright2020comparative} provide a comparative analysis of these guidelines.

Guidelines that focus on the design of AI systems, and specifically on the ethics of those systems, are critically important. Various attempts have been made to assist design practitioners in the process of operationalizing guidelines for AI systems, including guidebooks~\cite{pair2021people}, toolkits~\cite{deng2022exploring}, and checklists~\cite{madaio2020co}. When design guidelines are successfully applied, they make a positive impact, such as in assisting cross-functional development teams in improving user experiences~\cite{li2022assessing} and addressing ethical challenges~\cite{balasubramaniam2020ethical}. However, several studies have critiqued their comprehensiveness~\cite{hagendorff2020ethics}, the extent to which they can be operationalized~\cite{fukuda2021emerging}, and the lack of consequences when they are not followed~\cite{hagendorff2020ethics}. Additionally, \citet{madaio2020co} argues that the adoption of an AI ethics process within an organization, ``would only happen if leadership changed organizational culture to make AI fairness a priority, similar to priorities and associated organizational changes made by leadership to support security, accessibility, and privacy''~\cite[p. 8]{madaio2020co}.

Despite the preponderance of guidelines for human-AI interaction and AI ethics, there is a gap in the technologies on which they focus. To date, many of the AI guidelines developed within the HCI community primarily focus on discriminative AI~\cite{amershi2019guidelines, pair2021people, ibm2022design, ibm2022aiethics}, the class of algorithms that identifies boundaries that separate different classes or groups in a data set. These guidelines do not take into account generative AI algorithms that produce artifacts, rather than decision boundaries, as outputs. Because generative AI offers new ways for users to interact with technology, and raises new issues regarding the ethics of AI systems, a new set of design guidelines are needed.


\section{Why generative AI needs design principles}
\label{sec:why-genai-principles}

Generative AI technologies have introduced a new paradigm of human-computer interaction, what \citeauthor{nielsen2023ai} refers to as ``intent-based outcome specification''~\cite{nielsen2023ai}. In this paradigm, users specify \emph{what they want}, often using natural language\footnote{Some user interfaces to generative AI systems allow users to specify their intent via sketches and gestures~\cite{chung2021gestural}, UI controls~\cite{louie2020novice, liu2021neurosymbolic}, video-scanned body movements~\cite{wallace2021learning}, and even various forms of improv performance~\cite{beilharz2006hyper, jacob2013viewpoints}.}, but not \emph{how it should be produced.} One challenge of this paradigm stems from the distinguishing characteristic of generative AI: it generates artifacts as outputs and those outputs may vary in character or quality, even when a user's input does not change. This characteristic has been described by \citet{weisz2023toward} as \emph{generative variability,} and it provides what \citet{alvarado2018towards} describe as an ``algorithmic experience,'' raising questions on appropriate types of user control, levels of algorithmic transparency, and user awareness of how the algorithms work and how to effectively interact with them.

With generative AI applications, users will need to develop a new set of skills to work with (not against) generative variability by learning how to create specifications that result in artifacts that match their desired intent. One emerging skill revolves around crafting effective natural language prompts, known as in-context learning~\cite{dong2022survey, xie2022how, brown2020language} or prompt engineering~\cite{zamfirescu2023johnny, white2023prompt}. This process is typically informal and relies on trial-and-error~\cite{liu2022design, oppenlaender2022taxonomy, strobelt2022interactive, yang2023large}. The use of open-ended natural language, rather than a fixed vocabulary of commands, leads to new design challenges. For example, \citeauthor{nielsen1994enhancing} argues, ``users should not have to wonder whether different words, situations, or actions mean the same thing''~\cite[p.156]{nielsen1994enhancing}; given the innumerable ways that users can express their intent in a natural language prompt, how can generative AI applications help users achieve desired results? Is it necessarily a ``mistake'' or ``error'' when a user's prompt results in an output that they didn't anticipate or like? Does it violate the consistency heuristic when it is difficult for users to achieve replicable results (e.g.~\cite{megahed2023generative, perez2023risk, sharma2023generating}), because each click of the ``generate'' button results in different outputs, even for the same input?

Existing human-AI design guidelines fail to address the unique design challenges of generative AI because they do not cover generative use cases or new considerations stemming from generative variability, and they do not cover new or amplified ethical issues stemming from the models' generative nature. For example, guidelines published by \citeauthor{pair2021people} make recommendations such as, ``Design for labelers \& labeling'' and ``Design \& evaluate the reward function''~\cite{pair2021people}. The former of these recommendations will not apply to generative use cases that do not require data labeling, as the foundation models often used to implement generative capabilities are pre-trained and may not require additional labeled data for tuning. In addition, the latter recommendation is tailored to classification use cases in which false positives and false negatives are important outcome metrics; in generative use cases, these metrics have no meaning.

Guidelines from \citet{amershi2019guidelines} may be more readily adapted to generative AI applications, although their coverage of generative-specific considerations is limited and design practitioners may encounter difficulties in making such adaptations. For example, the recommendation to, ``Make clear why the system did what it did'' is potentially less important when a user's goal is to simply generate a desirable artifact\footnote{Research by \citet{sun2022investigating} explores the kinds of questions that users have when working with a generative AI system, which include questions about how an artifact was produced. We posit that the utility of a generated artifact need not depend upon the mechanics of \emph{how} that artifact was generated in the same way that a user's trust in a decision recommendation is often predicated on an explanation for how that recommendation was produced (e.g.~\cite{liao2020questioning, ashoori2019ai, vereschak2021evaluate}). Further, some applications of generative AI concern the exploration of a space of multiple possibilities (e.g.~\cite{kreminski2022evaluating, rost2023stable}), indicating that in some use cases, \emph{how} an artifact was generated may be of lesser importance than the generated artifacts themselves.}. The recommendation to, ``Make clear what the system can do'' may be difficult to implement in light of the emergent and unanticipated behaviors of generative foundation models~\cite{bommasani2021opportunities}, as well as the trial-and-error methods by which users iterate toward a desired outcome~\cite{liu2022design, oppenlaender2022taxonomy, strobelt2022interactive, yang2023large}.

Finally, alongside their tremendous potential to augment people's creative capabilities, generative technologies also introduce new risks and potential user harms. These risks include issues of copyright and intellectual property~\cite{lucchi2023chatgpt, franceschelli2022copyright}, the circumvention or reverse-engineering of prompts through attacks~\cite{deng2023jailbreaker}, the production of hateful, toxic, or profane language~\cite{hartvigsen2022toxigen}, the disclosure of sensitive or personal information~\cite{kim2023propile}, the production of malicious source code~\cite{chen2021evaluating, charan2023text}, and a lack of representation of minority groups due to underrepresentation in the training data~\cite{venkit2023nationality, garcia2023uncurated, luccioni2023stable, sun2019mitigating}. Work by \citet{houde2020business} takes concerns such as these to an extreme by envisioning realistic, malicious uses of generative AI technologies. Although it cannot be a designer's responsibility to curb all potentially-harmful usage, existing design guidelines for AI systems fall short in addressing these unique issues stemming from the \emph{generative} nature of generative AI, and AI ethics frameworks are only just starting to appear to provide designers with the language they need to begin discussing these important issues~\cite{weidinger2021ethical, ibm2022aiethics, deng2022exploring}.

We therefore conclude that there is a pressing need for a set of general design guidelines that help practitioners develop applications that utilize generative AI technologies in safer and more effective ways -- safer because of the new risks introduced by generative AI, and more effective because of the control that users have lost over the computational process. Although recent work has begun to probe at design considerations for generative AI, this work has been limited to specific application domains or technologies. For example, guidelines of various maturity levels exist for GAN-based interfaces~\cite{grabe2022towards, zheng2023stylegan}, image creation~\cite{liu2022design, liu2023beyond, verheijden2023collaborative}, prompt engineering~\cite{liu2022design, liu2023beyond}, virtual reality~\cite{urban2021designing}, collaborative storytelling~\cite{santiago2023rolling}, and workflows with co-creative systems~\cite{grabe2022towards, muller2020iccc, verheijden2023collaborative}.
Our work seeks to extend these studies toward principles that can be used \emph{across} generative AI domains and technologies.

\section{Design principles for generative AI applications}
\label{sec:design-principles}

\begin{table*}[htp]
    \small
    \centering
    \begin{tabularx}{\linewidth}{XX}
        \toprule
        \textbf{Design Responsibly} \newline Ensure the AI system solves real user issues and minimizes user harms 
        \begin{itemize}[leftmargin=*, nosep, after=\strut]
            \item \emph{Use a human-centered approach*}. Design for the user by understanding their needs and pain points, and not for the technology or its capabilities.
            \item \emph{Identify \& resolve value tensions*}. Consider and balance different values across people involved in the creation, adoption, and usage of the AI system.
            \item \emph{Expose or limit emergent behaviors}. Determine whether generative capabilities beyond the intended use case should be surfaced to the user or restricted.
            \item \emph{Test \& monitor for user harms*}. Identify relevant user harms (e.g. bias, toxic content, misinformation) and include mechanisms that test and monitor for them.
        \end{itemize}\nointerlineskip
        & 
        \textbf{Design for Generative Variability} \newline Help the user manage the ability of generative models to produce multiple outputs that are distinct and varied
        \begin{itemize}[leftmargin=*, nosep, after=\strut]
            \item \emph{Leverage multiple outputs}. Generate multiple outputs that are either hidden or visible to the user in order to increase the chance of producing one that fits their need.
            \item \emph{Visualize the user's journey}. Show the user the outputs they have created and guide them to new output possibilities.
            \item \emph{Enable curation \& annotation}. Design user-driven or automated mechanisms for organizing, labeling, filtering, and/or sorting outputs.
            \item \emph{Draw attention to differences or variations across outputs}. Help the user identify how outputs generated from the same prompt differ from each other.
        \end{itemize}\nointerlineskip
        \\
        \midrule
        \textbf{Design for Mental Models} \newline Communicate how to work effectively with the AI system, considering the user's background and goals
        \begin{itemize}[leftmargin=*, nosep, after=\strut]
            \item \emph{Orient the user to generative variability}. Help the user understand the AI system’s behavior and that it may produce multiple, varied outputs for the same input.
            \item \emph{Teach effective use}. Help the user learn how to effectively use the AI system by providing explanations of features and examples through in-context mechanisms and documentation.
            \item \emph{Understand the user's mental model*}. Build upon the user’s existing mental models and evaluate how they think about your application: its capabilities, limitations, and how to work with it effectively.
            \item \emph{Teach the AI system about the user}. Capture the user’s expectations, behaviors, and preferences to improve the AI system’s interactions with them.
        \end{itemize}\nointerlineskip
        &
        \textbf{Design for Co-Creation} \newline Enable the user to influence the generative process and work collaboratively with the AI system
        \begin{itemize}[leftmargin=*, nosep, after=\strut]
            \item \emph{Help the user craft effective outcome specifications}. Assist the user in prompting effectively to produce outputs that fit their needs.
            \item \emph{Provide generic input parameters}. Let the user control generic aspects of the generative process such as the number of outputs and the random seed used to produce those outputs.
            \item \emph{Provide controls relevant to the use case \& technology}. Let the user control parameters specific to their use case, domain, or the generative AI’s model architecture.
            \item \emph{Support co-editing of generated outputs}. Allow both the user and the AI system to improve generated outputs.
        \end{itemize}\nointerlineskip
        \\
        \midrule
        \textbf{Design for Appropriate Trust \& Reliance} \newline Help the user determine when they should or should not rely on the AI system’s outputs by teaching them to be skeptical of quality issues, inaccuracies, biases, underrepresentation, and other issues
        \begin{itemize}[leftmargin=*, nosep, after=\strut]
            \item \emph{Calibrate trust using explanations}. Be clear and upfront about how well the AI system performs different tasks by explaining its capabilities and limitations.
            \item \emph{Provide rationales for outputs}. Show the user why a particular output was generated by identifying the source materials used to generate it.
            \item \emph{Use friction to avoid overreliance}. Encourage the user to review and think critically about outputs by designing mechanisms that slow them down at key decision-making points.
            \item \emph{Signify the role of the AI}. Determine the role the AI system will take within the user's workflow.
        \end{itemize}\nointerlineskip
        &
        \textbf{Design for Imperfection} \newline Help the user understand and work with outputs that may not align with their expectations
        \begin{itemize}[leftmargin=*, nosep, after=\strut]
            \item \emph{Make uncertainty visible}. Caution the user that outputs may not align with their expectations and identify detectable uncertainties or flaws.
            \item \emph{Evaluate outputs using domain-specific metrics}. Help the user identify outputs that satisfy measurable quality criteria.
            \item \emph{Offer ways to improve outputs}. Provide ways for the user to fix flaws and improve output quality, such as editing, regenerating, or providing alternatives.
            \item \emph{Provide feedback mechanisms}. Collect user feedback to improve the training of the AI system.
        \end{itemize}\nointerlineskip
        \\
        \bottomrule
    \end{tabularx}
    \caption{Design \textbf{principles} and \emph{strategies} for generative AI applications. The left column contains principles that offer new interpretations of existing issues in the development of AI applications. The right column contains principles that focus on new issues that stem from generative AI technologies. Strategies that involve following a design process are indicated with an asterisk (*).}
    \label{tab:design_principles}
\end{table*}

We begin by presenting our final set of six design principles and their corresponding strategies in Table~\ref{tab:design_principles}, along with our overall design framework in Figure~\ref{fig:principles-schematic}. We also provide extended descriptions and examples of each principle and strategy in Appendix~\ref{appendix:extended-description}. In the rest of this paper, we describe the process we used to develop and validate these principles and strategies.

The principles are generally presented as high-level ``design for...'' statements that indicate the characteristics that are important to consider when making design decisions. Three principles focus on aspects of existing AI systems that have new interpretations through the lens of generative AI: \textbf{Design Responsibly}, \textbf{Design for Mental Models}, and \textbf{Design for Appropriate Trust \& Reliance}. Three principles identify unique aspects of generative AI UX: \textbf{Design for Generative Variability}, \textbf{Design for Co-Creation}, and \textbf{Design for Imperfection}.

Each design principle is coupled with a set of four design strategies for how to implement that principle. In some cases, implementing the principle involves following a design process; in other cases, it is implemented through the inclusion of specific types of features or functionality.

These principles and strategies can be employed to support two user goals: (1) \emph{optimization}, in which the user seeks to produce an output that satisfies some task-specific criteria; and (2) \emph{exploration}, in which the user uses the generative process to explore a domain, seek inspiration, and discover alternate possibilities in support of their own ideation. The ways each principle and strategy are applied may differ by user goal, and we elaborate on these differences in Section~\ref{sec:discuss-tasks}.

We note that these principles are just that -- \emph{principles} -- and not hard rules that must be followed in all design processes. Our view is that it is up to design practitioners to exercise their best judgement in deciding whether a principle applies to their particular use case, and whether any particular strategy should (or should not) be applied.

\section{Methodology}
\label{sec:methodology}

Our goal is to produce a set of clear, concise, and relevant design principles that can be readily applied by design practitioners in the design of applications that incorporate generative AI technologies. We aim for the principles to satisfy the following desiderata:

\begin{itemize}
    \item Provide designers with language to discuss UX issues unique to generative AI applications, motivated by work that provides designers with specialized vocabulary for domains such as video games~\cite{cairns2019future, khowaja2020framework} and IoT~\cite{chuang2018design};
    \item Provide designers with concrete strategies and examples that are useful for making difficult design decisions, such as those that involve trade-offs between model capabilities and user needs, motivated by work that focuses simultaneously on end-users of systems~\cite{gong2018establishment} and on designers as strategic and collaborative end-users of guidelines~\cite{kim2010effective, koelle2020social}; and
    \item Sensitize designers to the possible risks of generative AI applications and their potential to cause a variety of harms (inadvertent or intentional), and outline processes that could be used to avoid or mitigate those harms (e.g.~\cite{weidinger2021ethical, ibm2022aiethics}).
\end{itemize}

We used an iterative process to develop and refine the design principles, inspired by the process used by \citet{amershi2019guidelines} in developing their guidelines for human-AI interaction. We crafted an initial set of design principles via a literature search (Section~\ref{sec:iteration-1}), refined those principles via multiple feedback channels (Section~\ref{sec:iteration-2}), conducted a modified heuristic evaluation exercise to assess their clarity and relevance and identify any remaining gaps (Section~\ref{sec:iteration-3}), and finally applied the principles to two generative AI applications under design to demonstrate their applicability to design practice (Section~\ref{sec:iteration-4}).

\begin{figure*}
    \centering
    \includegraphics[width=\linewidth]{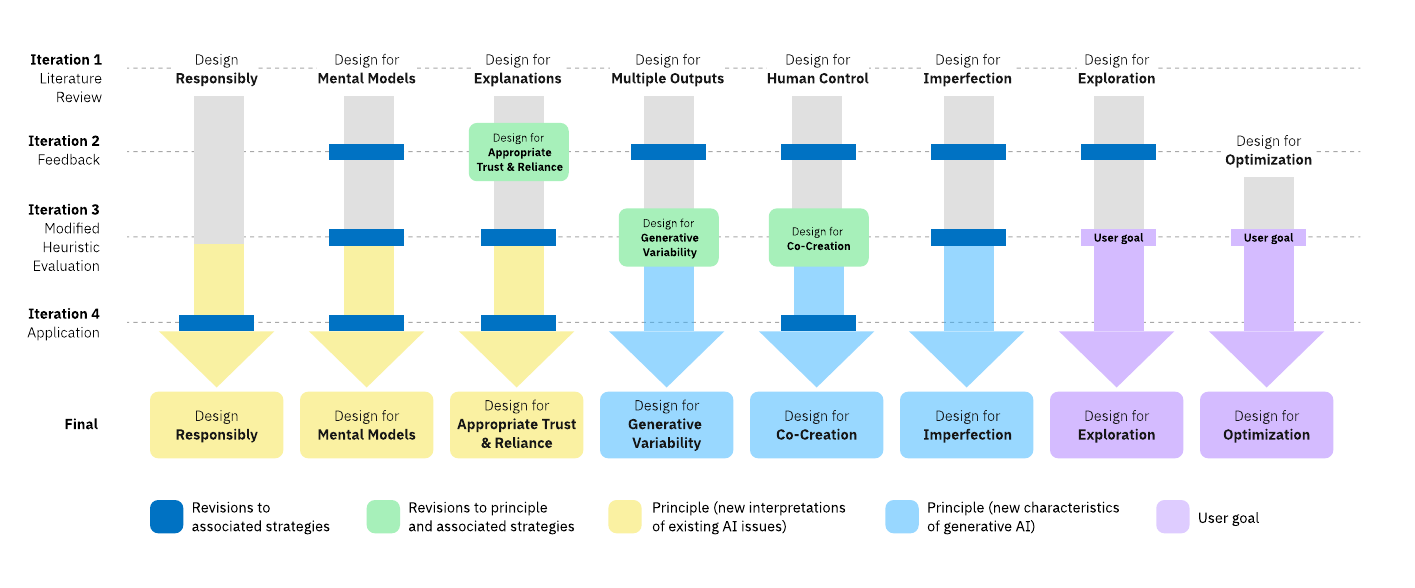}
    \caption{Evolution of the design principles across four iterations. During Iteration 3, we recognized that some principles offered new interpretations of existing AI system characteristics whereas others identified new characteristics of generative AI.}
    \label{fig:principle-evolution}
\end{figure*}

In each iteration, we engaged in significant discussion and reflection on the feedback gathered from the previous iteration to produce a new version of the design principles and strategies. In some cases, principles or strategies moved to the next iteration unchanged; in many cases, we made organizational and wording changes. We summarize our iterative process and the outcomes of each iteration in Table~\ref{tab:iterative-process} and we show how the principles evolved over the iterations in Figure~\ref{fig:principle-evolution}.

\begin{table*}[htp]
    \centering
    \begin{tabularx}{\linewidth}{lp{1.75cm}XX}
        \toprule
        \textbf{Iteration} & \textbf{Activity} & \textbf{Goal} & \textbf{Key Outcomes} \\
        \hline
        Iteration 1 & Literature\allowbreak Review & Identify relevant research and examples of generative AI application design & Observed hierarchy of high-level design principles implemented by specific UX design strategies; identified 7 initial design principles  \\
        \hline
        Iteration 2 & Feedback & Collect feedback from conference workshop and designers within our institution & Developed clearer understanding of Exploration vs. Optimization purposes of use \\
        \hline
        Iteration 3 & Modified Heuristic Evaluation & Test principles for clarity, relevance, and coverage by having designers evaluate commercial generative AI applications & Recognized uniqueness of Generative Variability, Co-Creation, and Imperfection to generative AI; re-categorized Exploration and Optimization separately as user goals \\
        \hline
        Iteration 4 & Application & Demonstrate real-world applicability and utility by having two product teams adopt the principles in their own design work & Observed utility of principles across ideation and evaluation design phases; made clarity improvements to strategy descriptions \\
        \bottomrule
    \end{tabularx}
    \caption{Summary of the iterative process we used to develop the design principles and strategies.}
    \label{tab:iterative-process}
\end{table*}

\section{Iteration 1: Crafting initial design principles}
\label{sec:iteration-1}

We began our process of identifying design guidelines suitable for generative AI applications by examining recent research in the HCI and AI communities. We conducted a literature review of research studies, guidelines, and analytic frameworks from these communities by searching the ACM Digital Library and Google Scholar for terms including ``generative AI, '' ``design guidelines,'' and ``human-centered AI.'' These searches identified a set of relevant publications, as well as several recent workshops covering human-AI interaction with generative AI: Human-AI Co-Creation with Generative Models~\cite{geyer2021hai, weisz2022hai, maher2023hai}, Generative AI and HCI~\cite{muller2022genaichi}, Human-Centered AI~\cite{muller2022hcai}, and Human-Centered Explainable AI~\cite{ehsan2022human}. We then conducted additional searches for terms found within those workshops' proceedings, including ``co-creation,'' ``human-AI collaboration,'' ``explainability,'' and ``creative interfaces.'' Our searches yielded a representative sample of work that included new advancements and issues in generative AI\footnote{Papers describing advancements and issues in generative AI, including demonstrations of new user interaction capabilities, included~\citet{perez2023risk, sharma2023generating, Brown:GPT3, johnson2007measuring, kaiser2018generative, louie2020novice, metz:GPT-3, ramesh2022hierarchical, rombach2022high, ross2023programmers, kim2023lmcanvas, liu2021neurosymbolic}.}, design guidelines\footnote{Papers offering design guidelines included~\citet{amershi2019guidelines, frijns2021design, fukuda2021emerging, jobin2019global, liu2022design, mohseni2021multidisciplinary, pair2021people, shneiderman2020bridging, srivastava2021actionable, urban2021designing, wickramasinghe2020trustworthy, wright2020comparative}.}, studies of design guideline implementation\footnote{Papers examining design guideline implementation included~\citet{alnanih2016mapping, li2022assessing, yildirim2022experienced}.}, and studies of human interaction with AI (and generative AI) systems\footnote{Papers examining human interactions with AI (and generative AI) systems included~\citet{gmeiner2023exploring, grabe2022towards, kreminski2022evaluating, lim2023generative, liu2023beyond, louie2020novice, maher2012computational, sun2022investigating, verheijden2023collaborative, wan2023felt, weidinger2021ethical, weisz2022better, ehsan2021expanding, liao2020questioning, megahed2023generative, weisz2021perfection, inie2023designing, deterding2017mixed, lubart2005can, muller2020iccc, seeber2020machines, spoto2017mici}.}. Finally, to incorporate recent industry developments around generative AI, we also examined a representative set of commercial generative applications (listed in Table~\ref{tab:genai-applications}) to identify common design patterns.

One characteristic that stood out to us in our review was the difference between work that identified important user needs and the specific kinds of UX design that supported those needs. For example, one set of papers examined requirements for explainable AI (XAI) and human-centered explainable AI (HCXAI) through experimental and heuristic methods~\cite{ehsan2021expanding, liao2020questioning, sun2022investigating}, motivating ``explainability'' as an important high-level concept. Then, when examining a commercial generative AI system (ChatGPT), we observed how explanations of the system's capabilities and limitations were provided on the home screen. Observations such as these motivated our development of a two-tier principle/strategy structure in which a principle articulates an important characteristic or consideration for a generative AI application and the strategies identify how to implement that principle in the UX.

Our analysis helped us identify several characteristics unique to generative AI that have implications on the user experience: the models' capability of producing multiple outputs~\cite{megahed2023generative, perez2023risk, sharma2023generating}, the possibility of flaws or imperfections\footnote{We identified two types of imperfection: (a) flaws present in the model's outputs, such as bugs in source code or hallucinations in Q\&A, and (b) misalignments between a user's intent, how that intent is expressed in a prompt, and what the generative model produces as a result. Even if category (a) would disappear (e.g. due to technological improvements), category (b) would remain and imperfection would still be a critical design issue.} within those outputs~\cite{weisz2021perfection, weisz2022better}, and the various ways that people can control or influence those outputs~\cite{kim2023lmcanvas, louie2020novice, liu2021neurosymbolic}. We also identified how generative AI could enable people to explore a space of possibilities~\cite{kreminski2022evaluating} as a byproduct of the generative process. In addition, we identified several existing considerations of AI systems as being particularly important to the generative case, such as using participatory methods~\cite{inie2023designing} to design for real user needs like explainability~\cite{sun2022investigating, ehsan2022human}, and understanding the role of the AI in the co-creative process~\cite{lubart2005can, deterding2017mixed, grabe2022towards, muller2020iccc, spoto2017mici, seeber2020machines}.

At this stage, we identified 7 high-level principles and 22 specific strategies for implementing them. Some strategies were related to multiple principles, and at this stage we allowed the overlap; in subsequent iterations, we eliminated these redundancies (we discuss this point further in Section~\ref{sec:guideline-organization}).


\section{Iteration 2: External and internal feedback}
\label{sec:iteration-2}

We published the first iteration of the design principles at the Human-AI Co-Creation with Generative Models (HAI-GEN) workshop at IUI~\cite{weisz2023toward}, attended by approximately 50 researchers from academia and industry. At this workshop, we received informal feedback through discussion sessions and follow-up conversations. We also published this version within our organization as part of a design guide on generative AI, which was viewed by over 1,000 design practitioners. We created an internal discussion channel on this guide to receive additional feedback, including points of confusion and gaps in our framework. Both sources of informal feedback helped us craft the second iteration of the design principles, which introduced the following major changes:

\begin{itemize}
    \item We identified how users' goals in using a generative AI system can differ, leading us to include two task-specific principles: the existing \textbf{Design for Exploration} principle, in support of use cases around ideation, exploration, and learning; and a new principle, \textbf{Design for Optimization}, in support of use cases for which the production of a singular artifact is desired.
    \item We recognized that explainability needs for generative AI systems, while important, were not necessarily an ``end'' in and of themselves. Rather, explainability is one way to \textbf{Design for Appropriate Trust \& Reliance}, leading us to incorporate existing explainability strategies into this new principle.
    \item We re-articulated all of the design strategies as rules of action (e.g. a verb followed by 2-6 words), akin to how \citeauthor{amershi2019guidelines} phrased their guidelines.
    \item We identified that five design strategies were about the design process itself rather than specific UX capabilities.
\end{itemize}

At the end of Iteration 2, we had a set of 8 high-level principles implemented by 29 specific strategies.


\section{Iteration 3: Modified heuristic evaluation}
\label{sec:iteration-3}

Following Iteration 2, we sought to conduct a more rigorous evaluation of the design principles and strategies. Given the potential gap between research literature and real-world practice, we specifically wanted to determine their clarity to our target audience of design practitioners, understand their relevance to commercial generative AI applications, and identify any additional gaps in our framework. In support of these goals, we drew inspiration from \citeauthor{amershi2019guidelines} by creating a modified heuristic evaluation exercise.

\subsection{Method}
Heuristic evaluation is a discount usability method for identifying violations of usability guidelines in a user interface~\cite{nielsen1990heuristic}. \citet{amershi2019guidelines} developed a \emph{modified} heuristic evaluation in which evaluators reviewed an AI-infused user experience with the purpose of evaluating the heuristics themselves. We similarly developed a modified heuristic evaluation to evaluate our design principles for generative AI applications. We asked evaluators to examine a range of commercial generative AI applications and identify examples that demonstrate the use of the principles and strategies, as well as examples of generative AI-specific design choices that were not covered by the principles and strategies. This exercise helped us evaluate the relevance, clarity, and coverage of the design principles and strategies.

We identified 9 commercial generative AI applications to use in the evaluation, listed in Table~\ref{tab:genai-applications}. We selected these applications due to their popularity in consumer or enterprise markets, their ability to be used within our organization without incurring costs, and the range of use cases and output modalities they supported. We also considered applications that incorporated generative AI features in one of two distinct ways\footnote{\citet{video:beavers2023} refer to different ``altitudes'' at which AI capabilities can be embedded in a UX, including ``above'' (the AI capability is the central focus of the application), ``beside'' (the AI capability sits in a side panel next to the main UI), and ``inside'' (the AI capability is embedded within the existing UI).}: either as the core user experience or as a component within an existing user experience.

\begin{table*}[ht]
    \centering
    \begin{tabularx}{\linewidth}{lXl}
        \toprule
        \textbf{Application} & \textbf{Description} & \textbf{AI Incorporation} \\
        \midrule
        ChatGPT & Conversational Q\&A & Core (Web app) \\
        Google Bard & Conversational Q\&A & Core (Web app) \\
        DALL-E & Text-to-image generator & Core (Web app) \\
        DreamStudio & Text-to-image generator & Core (Web app) \\
        Midjourney & Text-to-image generator & Component (Discord) \\
        Adobe Firefly Generative Fill & Text-to-image generator & Component (Adobe Photoshop) \\
        IBM watsonx.ai Prompt Lab & Prompt playground for large language models & Core (Web app) \\
        Github Copilot & Natural language to source code & Component (Visual Studio Code)\\
        AIVA & Music generation & Core (Web app) \\
        \bottomrule
    \end{tabularx}
    \caption{Commercial generative AI applications used in our modified heuristic evaluation. AI capabilities were present either as the core user experience or embedded as a component within an existing application.}
    \label{tab:genai-applications}
\end{table*}

We recruited 18 design practitioners within our organization and outside of our immediate team to perform the modified heuristic evaluation. We sought evaluators with varied design roles and levels of experience to ensure the principles were clear and relevant across different specialties and expertise levels. Of the 18 evaluators, 11 (61.1\%) identified as male, 6 (33.3\%) identified as female, and 1 preferred not to disclose. The majority of evaluators were User Experience Designers (16, 88.9\%), one evaluator was a Design Researcher, and one was a Research Software Developer\footnote{Despite not having a design role, this evaluator worked on an HCI research team and possessed over 20 years of professional design experience, which we felt sufficiently qualified them to participate in this exercise.}. Four evaluators (22.2\%) reported having 1-4 years of experience, four (22.2\%) had 5-9 years, two (11.1\%) had 10-14 years, two (11.1\%) had 15-19 years, and five (27.7\%) had 20+ years\footnote{As one evaluator did not complete the demographic survey, percentages do not add up to 100\%.}. Most evaluators had some experience with discount usability testing methods: three evaluators (16.6\%) reported low or very low experience, five (27.7\%) reported medium experience, and nine (50\%) reported high or very high experience. Given that generative AI design is an emerging field, our evaluators tended not to have high levels of experience in this area: eight evaluators (44.4\%) reported very low to low experience, eight (44.4\%) reported medium experience, and one (5.5\%) reported a high level of experience.

Evaluators self-selected an application familiar to them and completed their evaluation individually (as is standard practice~\cite{nielsen1990heuristic}) and remotely. As our evaluators were not involved in the design of these products, they were unable to evaluate the process-oriented strategies (all strategies within \textbf{Design Responsibly} plus \emph{Evaluate users' mental models}). Thus, these strategies were excluded from Iteration 3, and we made it a point to evaluate them in Iteration 4 (Section~\ref{sec:iteration-4}). Participants recorded their evaluations of all other principles in a Mural\footnote{Mural is a collaborative, graphical canvas application. https://mural.co} template. Two evaluators examined each application and each evaluation took approximately one hour.

We crafted short descriptions\footnote{These descriptions were the first iteration of those shown in Appendix~\ref{appendix:extended-description}.} for each principle and strategy to orient our evaluators. For each principle, we asked evaluators to begin by capturing examples in the Mural canvas of how their application applied the principle. At this stage, specific strategies in the Mural were covered with an overlay to encourage evaluators to find examples without being biased by our strategies, in hopes that they might identify new ones. After capturing examples, evaluators were instructed to remove the overlay, then label each example with a strategy we provided, ``not sure'', or a write-in for a new strategy. After finding and labeling examples, evaluators rated the relevance of each design principle and its strategies on a 4-point scale: ``Yes, they were clearly relevant,'' ``Yes, they were relevant but I struggled to find examples,'' ``No, they were clearly not relevant,'' and ``Not sure.'' They also rated the clarity of the principle as a whole on a 5-point scale from ``Very unclear'' to ``Very clear'' and provided suggestions for improvement. Finally, after reviewing all of the principles, evaluators were asked to identify any additional design features in their application that were not covered by the principles.

\subsection{Results}
\label{sec:iteration3-results}

Our evaluators produced 18 heuristic evaluation canvases laden with screenshots and sticky notes that identified real-world instances of the principles and strategies. They also left notes about points of difficulty or confusion.

\begin{figure*}
    \centering
    \includegraphics[width=\linewidth]{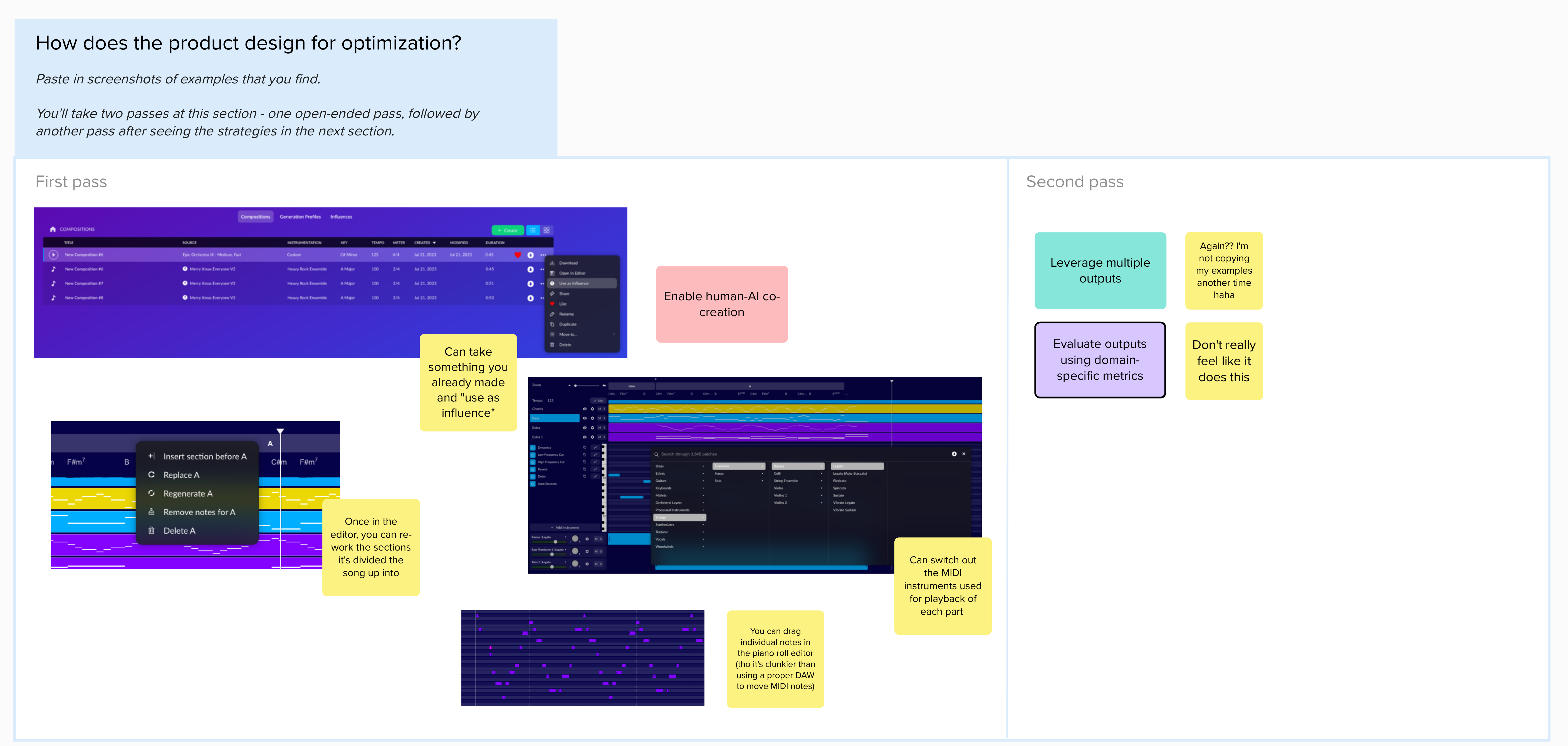}
    \caption{Portion of an evaluation of AIVA for the principle of \textbf{Design for Optimization}.}
    \label{fig:aiva-eval}
\end{figure*}

Figure~\ref{fig:aiva-eval} shows a portion of one evaluator's canvas in which they evaluated AIVA for \textbf{Design for Optimization}. This example shows how the evaluator found examples of various kinds of controls in the tool, along with feedback on the repetitiveness between \emph{Leverage multiple outputs}, \emph{Show multiple outputs}, and \textbf{Design for Multiple Outputs}: ``Again?? I'm not copying my examples another time.''

To analyze the evaluation data, two authors first individually examined the completed canvases for examples and comments that indicated the relevance, clarity, and coverage of the principles. They also reviewed participants' ratings of relevance and clarity and delved into their examples and comments to understand instances of lower ratings. They then converged with the other authors to review their findings and discuss potential ways to improve the principles and strategies.

\subsubsection{Relevance}
\label{sec:he-relevance}

\begin{figure*}
    \centering
    \includegraphics[width=\linewidth]{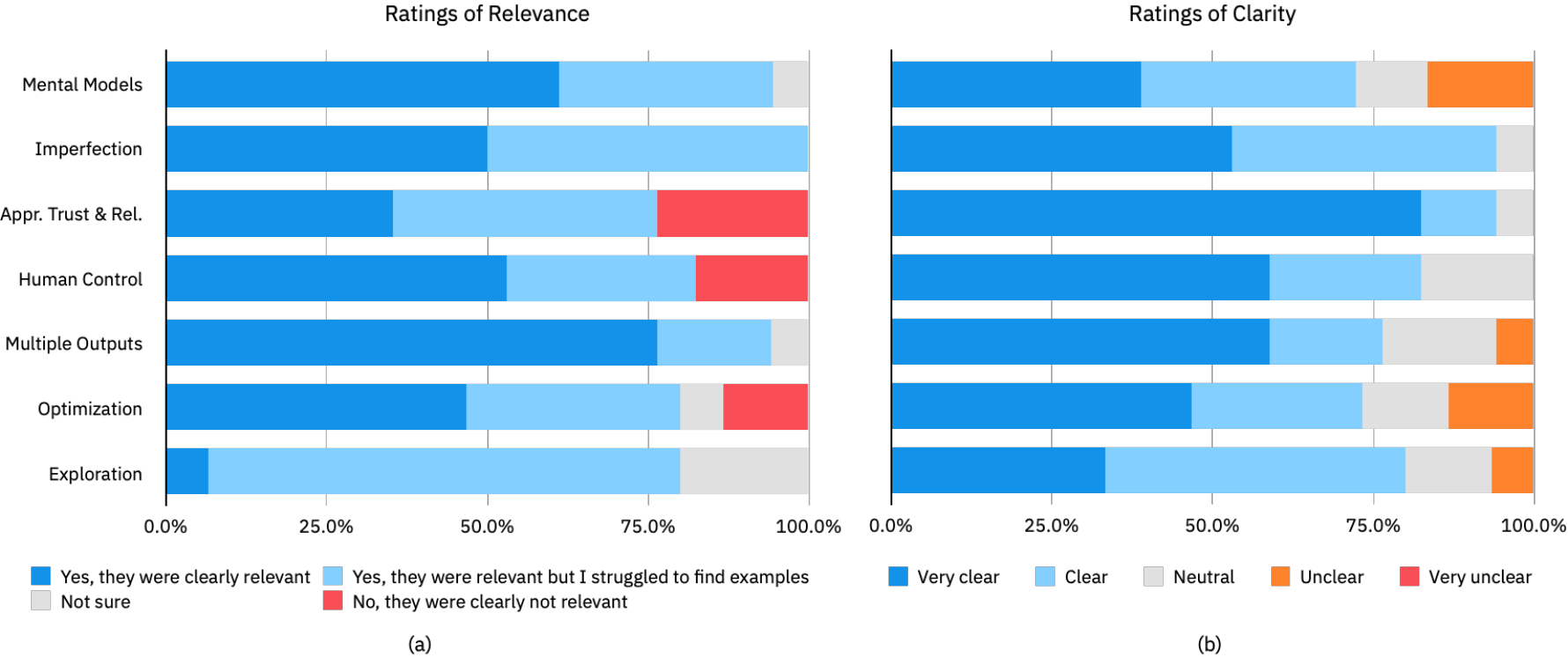}
    \caption{Evaluators' ratings of the (a) relevance and (b) clarity of each principle and its strategies to their application in the modified heuristic evaluation.}
    \label{fig:relevance-clarity-ratings}
\end{figure*}

To assess the relevance of the design principles and strategies to commercial generative AI applications, we counted the number of examples evaluators found. Evaluators identified 286 total examples across all principles; they found a collective average of 11.9 examples for each strategy, and every strategy had at least one example. The wealth of examples found suggests the principles and strategies were relevant to a range of commercial generative AI applications. Evaluators also generally rated each principle as being relevant (Figure~\ref{fig:relevance-clarity-ratings}a).

The relatively lower relevance ratings for \textbf{Design for Appropriate Trust \& Reliance}, \textbf{Design for Human Control}, and \textbf{Design for Optimization} stemmed from differences in application domain and output modality. In some cases, we accepted that relevance may vary by use case; in other cases, we addressed issues raised by participants to clarify or expand relevance. For example, the four evaluators who rated \textbf{Design for Appropriate Trust \& Reliance} as ``not relevant'' had examined image or music generation applications and felt that overreliance was less of a concern for creative applications. In response to this observation, we added examples of risks to be wary of in creative outputs (e.g. quality issues, bias, and underrepresentation~\cite{ibm2023foundation, bird2023typology, edenberg2023disambiguating}) to clarify its relevance to such applications. We made similar modifications to strategies that were too narrowly focused on specific domains or output modalities.

\subsubsection{Clarity}
\label{sec:he-clarity}

To assess the clarity of the design principles and strategies, we identified instances where evaluators noted overlap or redundancy between different principles or strategies, expressed confusion, or interpreted a principle or strategy differently from how we intended. We also asked evaluators to rate the clarity of each principle (Figure~\ref{fig:relevance-clarity-ratings}b), and they were generally rated as being clear.

Participants identified eight overlap issues. Notably, five evaluators found that nearly all strategies in \textbf{Design for Exploration} and \textbf{Design for Optimization} overlapped in some way with strategies in other principles. This observation led us to reconsider how to incorporate exploration and optimization within our framework. Ultimately, we recognized that exploration and optimization are user goals rather than characteristics of a generative AI application, and hence should be communicated as such (we discuss this point further in Section~\ref{sec:discuss-tasks}). Other overlap issues were reconciled by merging redundant strategies.

We observed 16 instances in which an evaluator's use of a strategy label mismatched our intention for what the strategy represented. We made two major changes in response to these mismatches. First, we reframed \textbf{Design for Human Control} as \textbf{Design for Co-Creation} in response to frequent misinterpretations of ``controls'' as affordances unrelated to the generative process (such as Photoshop's editing tools). \textbf{Design for Co-Creation} provides greater specificity to generative AI's unique capabilities for human-AI co-creation, which has been examined extensively within HCI communities (e.g.,~\cite{davis2016empirically, muller2023interactional, oliveira2014adapting, gluaveanu2014distributed, kantosalo2020five}). The second change was to rename \textbf{Design for Multiple Outputs} to \textbf{Design for Generative Variability} to better characterize its purpose after observing that many evaluators narrowly interpreted this principle as solely being about the \emph{display} of multiple outputs. We made additional wording changes and clarifications to other principles and their associated strategies in response to participants' feedback.

\subsubsection{Coverage}
\label{sec:he-coverage}

Evaluators found new examples that reflected gaps in our framework, resulting in three new strategies: \emph{Teach the AI system about the user}, \emph{Help the user craft effective outcome specifications}, and \emph{Support co-editing of generated outputs}. We included \emph{Teach the AI system about the user} in \textbf{Design for Mental Models} as it addresses recent research in Mutual Theory of Mind~\cite{cuzzolin2020knowing, wang2021towards, winfield2018experiments}. We included \emph{Help the user craft effective outcome specifications} and \emph{Support co-editing of generated outputs} in \textbf{Design for Co-Creation} as they are most closely related to the co-creative process.


\section{Iteration 4: Application to generative AI UX design}
\label{sec:iteration-4}

Design guidelines can be difficult to put into practice~\cite{hagendorff2020ethics, mariage2011state, soni2019framework, stark2016bridging,
wang2023designing, yildirim2023investigating}, often because they describe goals rather than actions~\cite{johnson2020designing}. Our strategies were meant to capture ``actions'' that practitioners could take to apply the principles to their work. After refining the principles and strategies for relevance and clarity, we evaluated their utility within the design process by conducting structured, exploratory workshops with design practitioners within our organization who work on generative AI applications. Our primary goal was to understand how effectively the design principles could be applied in practice, but we remained open to identifying additional issues regarding relevance, clarity, and coverage.

\subsection{Method}

We held two workshops with two different teams (Table~\ref{tab:workshop-participants}) to evaluate the design principles and strategies in practice. Workshop 1 was held with an internal team comprised of four design practitioners working on the IBM watsonx.ai Prompt Lab\footnote{Watsonx.ai. \url{https://watsonx.ai}}, a prompt testing environment for large language models. Workshop 2 involved a separate internal team of ten design practitioners in the early, formative stages of designing an internal LLM-based conversational tool that provides UX research support. We selected these two teams as they provided a broader view on the actionability of the design principles in different phases of design: a later, evaluative stage (Workshop 1) and an earlier, ideation phase (Workshop 2).

\begin{table}[htp]
    \centering
    \begin{tabularx}{\linewidth}{lXlXlX}
        \toprule
        \multicolumn{2}{X}{\textbf{Workshop 1}} & \multicolumn{2}{X}{\textbf{Workshop 2}} \\
        \multicolumn{2}{X}
        {Evaluate existing LLM prompt tool} & \multicolumn{2}{X}{Ideate on a future LLM-based conversational tool} \\
        \midrule
        P1-1 & Design Lead       & P2-1 & UX Researcher   & P2-6  & Design Research Lead\\
        P1-2 & Software Designer & P2-2 & UX Researcher   & P2-7  & Design Manager\\
        P1-3 & Design Manager    & P2-3 & UX Researcher   & P2-8  & UX Researcher \\
        P1-4 & UX Researcher     & P2-4 & Program Manager & P2-9  & Technical Program Lead\\
             &                   & P2-5 & UX Researcher   & P2-10 & Design Research Lead \\
        \bottomrule
    \end{tabularx}
    \caption{Participants in each of the workshops to assess the actionability of the design principles.}
    \label{tab:workshop-participants}
\end{table}

Workshops took place remotely via video conferencing and were recorded with participants' consent. Each session lasted 90 minutes and included two moderators and two note-takers. One moderator began each workshop by presenting an overview of the design principles and strategies. To minimize the time required of participants, we split each session into two groups and assigned three principles to each group. Each break-out group contained one moderator and one note-taker.

For each principle, participants were first asked to identify ways they were already ``designing for'' or considering the principle. Next, they identified relevant strategies that they had not yet considered and brainstormed ways to leverage them to improve their product. This brainstorming session produced new design ideas that team members shared and discussed with each other. At the end of the session, participants reflected on the actionability of the principles within their design process. The moderators and note-takers of each workshop reviewed participants' design ideas and recording transcripts to identify insights on the usefulness of the principles and recommendations for improvement.

\subsection{Results}

\subsubsection{Applicability to practice}

Participants in Workshop 1 brainstormed a total of 46 design ideas and participants in Workshop 2 brainstormed 56 design ideas. These design ideas included feature requirements, new affordances, design processes to try, and questions to consider when making design decisions. Groups generated between 5 and 14 ideas per principle, and participants generated multiple, varied ideas for all of the principles. For example, when considering the strategy, \textit{Help the user craft effective outcome specifications}, P1-3 thought of an idea to ``provide different `effects' that bake in some prompt content, (e.g. in the style of a famous author),'' and P1-4 proposed a system to ``reward `best in class' prompt authors and celebrate and share'' their work.

When asked about the actionability of their brainstormed ideas, P1-1 responded, ``I definitely think a lot of these could go on a future roadmap.'' P2-1 commented that the workshop, ``made it clear crucial blind spots that could put the [application] idea at risk if not addressed,'' and that it helped their team, ``quickly generate new requirements.'' Participants' breadth of design ideas and comments on workshop outcomes indicate that practitioners are able to leverage the principles and strategies to inform useful, actionable design improvements.

Participants also shared ideas on how to improve the actionability of the principles. As their understanding of the principles was limited to the brief overview provided at the start of the workshop, they felt that having more details and resources to learn about the principles and strategies would make them easier to apply. P1-3 commented, ``having some examples of these concepts out in the world or in other tools might be a useful way to get a grasp of how the concept works.'' In support of this need, we include a library of examples in Appendix~\ref{appendix:extended-description} that provide richer detail on how each strategy has been applied within existing applications.

Participants also shared insights on how the principles would be incorporated into their design process. P1-4 commented that involving more roles in a workshop, such as developers and product managers, would add value. P2-8 felt that \textbf{Design for Imperfection} ``can't be applied unless user research is done'' due to a lack of understanding of the user's expectations for model outputs. P2-6 also spoke about the value of user research since ``the idea or solution... may look differently depending on the user.'' These comments indicate that a baseline understanding of users is needed to identify concrete design ideas from the principles and strategies, in line with our recommendation to \emph{Use a human-centered approach}. 

The outcomes from these workshops demonstrated that the design principles can be applied in both an early ideation stage and a later evaluation stage to drive actionable design ideas.

\subsubsection{Relevance and clarity improvements}

Workshop participants also provided feedback on the relevance and clarity of the design principles. This feedback primarily resulted in minor wording changes to the process-oriented strategies that we were unable to evaluate in Iteration 3; no major organizational changes were made as a result of this feedback. After incorporating this feedback, we produced our final set of design principles and strategies (Table~\ref{tab:design_principles}).

\subsection{Final clarity evaluation}

To determine whether the wording changes we made after Iterations 3 \& 4 impacted the clarity of the final design principles and strategies, we ran a follow-up survey with the evaluators from Iteration 3. Fourteen of 18 evaluators responded to our survey (77.7\% response rate).

The clarity ratings collected in Iteration 3 were at the higher end of the 5-point scale (M (SD) = 4.29 (0.90) of 5). The changes we made in Iteration 4 made a small but significant improvement to clarity (M (SD) = 4.58 (0.58) of 5), $F(1, 184) = 5.88$, $p = .02$, $\eta_{p}^{2} = .03$ (small).


\section{Discussion}
\label{sec:discussion}

We identified a set of six principles important to the design of generative AI applications, along with a companion set of 24 strategies for implementing those principles within a user experience. These principles were developed iteratively using a combination of critical conceptual analyses (to ensure scientific validity) and empirical work (to ensure real-world utility). 

We collected formal feedback on the principles from a 18 design practitioners who collectively evaluated them against 9 commercial applications. We then collected feedback from 12 design practitioners on two design teams who applied them to both the formative and evaluative stages of product design. We found that the principles helped design practitioners generate useful and actionable design improvements and were applicable to a range of generative AI applications, including those that generate different types of media (e.g. text, images, music).

We discuss two issues that kept surfacing throughout our development process that required us to think deeply about where to ``draw the lines,'' either between different principles and their strategies when we identified overlap or redundancy, or between what we later identified as a difference between user goals and characteristics of generative AI. We also discuss our strategies for putting the principles into action within our organization, as well as limitations and opportunities for future work.

\subsection{Guideline organization}
\label{sec:guideline-organization}

Early in our first iteration, we observed a hierarchical relationship emerge between high-level design principles that identified unique or differentiated aspects of generative AI and lower-level strategies for implementing those principles in a user experience. However, the relationships between which strategies applied to which principles were not always clear, as some strategies could be used to support multiple principles. For example, in Iteration 1, the strategy \emph{Visualizing differences} was included in \textbf{Design for Multiple Outputs}, as it could help users understand the differences amongst those outputs (especially for use cases where those differences might be subtle). But it was also included in \textbf{Design for Imperfection}, as it could help users more easily identify problematic outputs. Another example from Iteration 1 was the use of a \emph{Sandbox / Playground Environment}, which supported \textbf{Design for Imperfection} by not tainting an artifact-under-creation with potentially-problematic generated content (e.g. source code with bugs or text containing factual errors). But it was also included in \textbf{Design for Exploration}, as a sandbox provides a separate space for users to explore new candidates without interfering with their main working environment.

As we worked through subsequent iterations, we wrestled with whether we should continue to allow strategies to overlap between principles or aim for a clean separation. During Iteration 2, with the delineation between \textbf{Design for Exploration} and \textbf{Design for Optimization}, even more redundancy was introduced as many strategies support both kinds of uses. At this point, we even considered completely decoupling the strategies from the principles and providing an indication on each strategy (such as a tag) for which principle(s) it supported.

We ultimately decided to maintain the nesting of strategies within principles and aim for establishing clean boundaries. We made this decision because the amount of overlap diminished when we refined the principles during Iteration 3 and separated out the user goals of optimization and exploration. Our evaluators also experienced frustration when they were unable to differentiate between strategies, indicating a need to eliminate overlaps. However, we note that a single UX feature may be used to implement more than one principle or strategy (see Appendix~\ref{appendix:extended-description} for examples); therefore, we only sought to reduce conceptual overlaps between the principles and strategies themselves, as opposed to overlaps when a specific UX capability addresses multiple principles or strategies.

\subsection{User goals versus design principles}
\label{sec:discuss-tasks}

In reviewing the \emph{Library of Mixed Initiative Creative Interfaces}~\cite{spoto2017mici}, we realized that generative capabilities are sometimes an end in themselves, but other times are a means to achieving another goal. We identified these two different purposes of use as optimization and exploration, respectively:

\begin{itemize}
    \item In optimization use cases, the process of generating artifacts is an end: users use the generative capability to produce one or more artifacts that satisfy their needs, such as a source code function that implements a desired operation, a molecule that possesses specific properties, or an image that depicts a desired scene or character. We labeled this class of usage as ``optimization'' in recognition that the generative AI model may not produce a flawless or ``perfect'' output, and some amount of refinement (either by the user or the AI) may be required before it is satisfactory.
    \item In exploratory use cases, the process of generating artifacts is a means to an end: the purpose is not to generate the artifact, but to use the generated artifacts in order to learn about a domain (e.g. programming~\cite{liffiton2023codehelp} or medicine~\cite{kung2023performance}) or be inspired by seeing new or different possibilities (e.g. brainstorming~\cite{salikutluk2023interacting} or pre-writing~\cite{wan2023felt}). Few other types of AI technology support this kind of usage, where the emphasis is on assisting people in conducting a thought process.
\end{itemize}

During Iteration 3, we ultimately decided to remove \textbf{Design for Exploration} and \textbf{Design for Optimization} as core design principles because of the strong degree of overlap between their strategies and the strategies of other principles. In fact, we could not even clearly delineate different generative AI applications as supporting exploratory versus optimization usage, because many applications supported both, and users might even alternate between the two kinds of usage when using the application. For example, work by \citet{weisz2022better} shows how software engineers used generative technologies not only to produce source code translations (optimization), but also to improve their own knowledge of programming (exploration), within the same overall task context. Hence, we drew a line between user goals and design principles (depicted in Figure~\ref{fig:principles-schematic}).

We assert that each principle broadly supports both user goals, but the extent of their support does differ by goal. \textbf{Design for Imperfection} is strongly aligned with optimization use cases, as the reason why optimization is even necessary is because of the imperfect outputs produced by generative models. Concurrently, \textbf{Design for Generative Variability} is strongly aligned with exploration use cases, as generative variability is a key enabler of exploration. However, we note that \textbf{Design for Imperfection} can also support exploratory use by embracing unexpected ``imperfections'' that arise from discrepancies between a user's intent and the model's output. In addition, \textbf{Design for Generative Variability} can also support optimization use cases by helping users narrow down on an option that fits their needs from a wide field.

Another principle that has a high degree of affinity to optimization is \textbf{Design for Appropriate Trust \& Reliance}, especially when generated outputs are used within high-stakes domains (e.g. code, customer service). Trust and reliance may be of a lesser concern for exploration use cases, although users should still be wary of bias, underrepresentation, and other harms that may occur.

Finally, \textbf{Design for Co-Creation} can be applied to both exploration and optimization tasks, but the strategies of \emph{Help the user craft effective outcome specifications} and \emph{Support co-editing of generated outputs} may be more important when users need to optimize generated outputs to fit certain criteria.

We conclude that the principles and strategies form a toolbox that design practitioners can use holistically or selectively as they craft user experiences for generative AI applications. Design practitioners know their users and their needs best -- exemplified by \emph{Use a human-centered approach} -- and it is our hope that we have provided useful vocabulary for them to understand and design for the new and different kinds of uses that generative AI systems offer.

\subsection{Adoption within our organization}

As discussed in Section~\ref{sec:guidelines-hci}, the HCI community has produced a prodigious number of design guidelines throughout its history. But, as noted by both \citet{soni2019framework} and \citet{stark2016bridging}, our community struggles with bridging the gap between the development of scientifically-grounded guidelines and real-world design practice. We developed our design principles specifically to provide practical and actionable support to design practitioners. Therefore, we undertook a number of efforts to promote their adoption within our organization.

\begin{enumerate}
    \item \textbf{Actionable activities}. To bridge the gap between theory and practice, we developed activities for designers to apply the principles and strategies to their own work. Chief among them is a heuristic evaluation that uses the principles as heuristics for designers to evaluate the user experience of generative AI applications. We created and disseminated a self-contained Mural template that guides designers through this evaluation to identify new ideas and opportunities for design improvement. We also developed workshop activities for identifying applications of generative AI that drive user value and evaluating a user's mental model of an AI system.
    
    \item \textbf{Progressive detail}. When we initially developed the principles and strategies, we wrote about them extensively in a comprehensive guide that provided foundational knowledge and case studies on generative AI, which we shared with our internal design community. We received feedback that the level of detail was informative but too lengthy for busy designers. In response, we developed two condensed presentations: 1) paragraph-length descriptions for each principle and strategy (shown in Appendix~\ref{appendix:extended-description}) which were included in the generative AI heuristic evaluation template, and 2) one-sentence descriptions of each principle and strategy (shown in Table~\ref{tab:design_principles}) which were published on an internal website for the design of AI applications.

    \item \textbf{Hands-on outreach}. We conducted outreach activities to raise awareness of the principles within our organization. Some of these efforts targeted a general design audience, such as creating a discussion group\footnote{This group grew to over 1,200 members over the course of 9 months.} for generative AI design and presenting the principles at internal seminars. Other outreach targeted designers on key product teams. As one example, we held a workshop attended by 62 people at an internal design event to teach designers how to conduct a heuristic evaluation of generative AI applications. Instead of using a sample application, we evaluated a product recently released by our organization and invited the product's design team to participate. In an hour-long session, participants identified 10 usability issues and 6 new feature ideas, which we discussed in detail with the product team in follow-up meetings. They reported our findings to be useful and included several recommendations in their roadmap.

    \item \textbf{Executive sponsorship}. In addition to bottom-up dissemination, we also worked with key executives in our design organization to encourage relevant product teams to adopt the principles (as recommended by~\citet{madaio2020co}). Through this effort, we introduced the principles to 10 product teams who were in the process of learning about generative AI and identifying opportunities for incorporating it into their product.
\end{enumerate}

Akin to \citeauthor{yildirim2023investigating}'s observations on how their guidebook improved AI literacy within their organization and helped designers establish credibility and advocate for user needs~\cite{yildirim2023investigating}, we found our materials had a similar impact. The executive sponsorship of our work and the adoption of the principles by numerous product teams speak not just to their practical utility, but also for the great need to equip design practitioners and enable them to ``have a seat at the table'' in the creation of generative AI applications.

\subsection{Limitations and future work}

The field of generative AI is undergoing rapid innovation, both in the pace of technological development and in how those technologies are being brought to the market. We view our principles as beginning a discussion on how to design effective and safe generative AI applications. As the pace of innovation continues and new generative AI applications are developed, we anticipate new challenges to be uncovered, necessitating new sets of guidelines, tools, best practices, design patterns, and evaluative methods.

One challenge we encountered with the modified heuristic evaluation was in its use to evaluate \emph{the design principles themselves} through the process of evaluating a generative AI application. Not all of our evaluators understood this distinction, and as a result, we sometimes received feedback about shortcomings of the applications that was less relevant to our goal of improving the principles. We recommend providing stronger introductory examples that focus on how they help evaluate the principles rather than the products.

Another limitation of the modified heuristic evaluation was our focus on evaluating commercially-available generative AI applications. There are also many experimental applications in this space, but we did not examine them. Our restriction to commercial applications excluded other ways of interacting with generative AI applications, such as through narrative~\cite{calderwood2020novelists}, lyric and other poetic forms~\cite{schober2022passing}, and movement~\cite{wallace2021learning}.

Finally, our design guidelines are entirely focused on helping design practitioners to develop the user experience for a generative AI application. But, UX design is only one portion of the AI development lifecycle, which includes other phases such as model selection, model tuning, prompt engineering, deployment \& monitoring, and more. As decisions made during those phases will ultimately impact the user experience, we believe design practitioners ought to have their inputs considered. However, the design principles do not currently help them understand, for example, how to determine which generative model should be used to implement a Q\&A use case and when that model's performance is ``good enough,'' or how to hide a generative model's inference latency. In addition, organizational policies may be created that govern the uses of generative AI. We believe there is room for expansion to identify how designers can participate in these kinds of technical and policy decisions that have an impact on the user experience.


\section{Conclusion}

We developed a set of six principles for the design of applications that incorporate generative AI technologies. Three principles -- \textbf{Design Responsibly}, \textbf{Design for Mental Models}, and \textbf{Design for Appropriate Trust \& Reliance} -- offer new interpretations of known issues with the design of AI systems when viewed through the lens of generative AI. Three principles -- \textbf{Design for Generative Variability}, \textbf{Design for Co-Creation}, and \textbf{Design for Imperfection} -- identify issues that are unique to generative AI applications. Each principle is coupled with a set of strategies for how to implement it within a user experience, either through the inclusion of specific types of UX features or by following a specific design process. We developed the principles and strategies using an iterative process that involved reviewing relevant literature in human-AI collaboration and co-creation, collecting feedback from design practitioners, and validating the principles against real-world generative AI applications. We also demonstrated the value and applicability of the principles by applying them in the design process of two generative AI applications. As generative AI technologies are rapidly being incorporated into existing applications, and entirely new products are being created with these technologies, we see significant value in principles that aid design practitioners in harnessing these technologies for the benefit of their users in safe and effective ways.

\begin{acks}
    We thank everyone at IBM who provided valuable feedback and guidance in the development of these design principles.
\end{acks}

\bibliographystyle{ACM-Reference-Format}
\bibliography{references}

\appendix
\section{Extended descriptions and examples}
\label{appendix:extended-description}

We provide extended descriptions for each design principle and strategy to offer deeper insight into their meaning and application, written in second-person for an audience of design practitioners. We also provide examples for each design strategy to illustrate how they have been used in a realistic context. Each example was drawn either from a commercial generative AI application or an experimental generative AI system. Most examples were identified in the modified heuristic evaluation of Iteration 3; new examples were found for the three new strategies added after that iteration. For process-related strategies (denoted with an asterisk (*)), we discuss how the process would theoretically be applied as we did not have visibility into the actual design process for the generative AI applications we examined.

We make reference to the following commercial systems in our examples:

\begin{itemize}
    \item Adobe Firefly: \url{https://firefly.adobe.com}
    \item Adobe Photoshop: \url{https://www.adobe.com/products/photoshop.html}
    \item AIVA: \url{https://www.aiva.ai/}
    \item ChatGPT: \url{https://chat.openai.com}
    \item DALL-E: \url{https://labs.openai.com}
    \item DreamStudio (powered by Stable Diffusion): \url{https://dreamstudio.ai}
    \item Github Copilot: \url{https://github.com/features/copilot/}
    \item Google Bard: \url{https://bard.google.com}
    \item Midjourney: \url{https://midjourney.com}
\end{itemize}

We also note that a single type of UX feature or functionality can be used to implement more than one design strategy. We have included similar or duplicate examples below to illustrate this point.

\subsection{Design responsibly}
The most important principle to follow when designing generative AI systems is to design responsibly. The use of all AI systems, including those that incorporate generative capabilities, may unfortunately lead to diverse forms of harms, especially for people in vulnerable situations. As designers, it is imperative that we adopt a socio-technical perspective toward designing responsibly: when technologists recommend new technical mechanisms to incorporate into a generative AI system, we should question how those mechanisms will improve the user’s experience, provide them with new capabilities, or address their pain points.

\subsubsection{Use a human-centered approach*}
Technosolutionism is the idea that technology will solve all of our (human) problems, and it should be avoided at all costs. Human-centered approaches can determine whether the use of generative AI is appropriate; use cases that apply these technologies for their own sake may fail to deliver real user value.

\textit{Example}: Human-centered approaches such as design thinking and participatory design allow you to observe users' workflows and pain points to ensure proposed uses of generative AI are aligned with users' actual needs. For example, involving stakeholders in the co-design of prototypes can serve as a probe for discussions around user value and technological feasibility.

\subsubsection{Identify and resolve value tensions*}
Often, there are multiple stakeholders involved in the creation of a generative AI application, including the end users, those who design and build the application (e.g. designers, developers, product managers), and those who make purchasing or licensing decisions (e.g. CIOs, CEOs). When these stakeholders' values are not aligned, it results in a value tension. Addressing these tensions is important for creating feasible and valuable products that meet end users' needs.

\textit{Example}: Value Sensitive Design (VSD)~\cite{friedman1996value} is a method that can help designers identify who the important stakeholders are and navigate the value tensions that exist across them.

\subsubsection{Expose or limit emergent behaviors*}
Generative AI can exhibit emergent behaviors — the ability to perform tasks beyond ones they were trained for. These emergent behaviors can be a delighter, such as when a conversational Q\&A system answers an out-of-domain question, or they can be a risk, such as when that output is toxic or aggressive. As a designer, you should carefully consider the trade-offs between optimizing your user experience for a well-defined set of capabilities versus providing a more open-ended experience that may surface potentially risky emergent behaviors.

\textit{Example}: Conversational interfaces that enable open-ended interactions will allow such emergent behaviors to surface. For example, a user may discover that ChatGPT can perform sentiment analysis, a task that it (likely) wasn't explicitly trained to do. By contrast, graphical user interfaces (GUIs), such as AIVA, can place limits on the ways a user can interact with the underlying generative model by only exposing selected functionality.

\subsubsection{Test \& monitor for user harms*}
Generative models may produce a variety of harmful outputs, such as language that is hateful, abusive, profane, or otherwise toxic. They may also harm users by failing to produce outputs that provide a fair or accurate representation of diversity. It is imperative to work closely with technologists to understand and evaluate the potential risks stemming from the use of a generative model in an application. It is also imperative to assume that user harms will occur and develop reporting and escalation mechanisms for when they do.

\textit{Example}: One way to test for harms is by benchmarking models on known data sets of hate speech~\cite{hartvigsen2022toxigen} and bias~\cite{felkner2023winoqueer, shaikh2022second, venkit2023nationality}. After deploying an application, harms can be flagged through mechanisms that allow users to report problematic model outputs.

\subsection{Design for mental models}
A mental model is a simplified representation of the world that people use to process new information and make predictions~\cite{kelly2023capturing}. It is their own understanding of how something works and how their actions affect it. Generative AI poses new challenges to users, and designers must carefully consider how to impart useful mental models to their users to help them understand how a system works and how their actions affect that system. Also consider the user’s background and goals and how to help the AI form a ``mental model'' of the user.

\subsubsection{Orient the user to generative variability}
Help the user understand the AI system’s behavior, and that it may produce multiple, varied outputs that may not be reproducible, even when given the same input. This behavior will be unexpected for novice users because it is fundamentally different from traditional AI systems that always give the same outcome for the same input.

\textit{Example}: Google Bard provides answers in the form of multiple drafts, indicating that it came up with multiple, varied answers for the same question.

\subsubsection{Teach effective use}
Users need to understand how to work effectively with a generative AI application to accomplish their goals. Mechanisms such as tutorials, examples, explanations, and social transparency~\cite{ehsan2021expanding} (i.e. showing other users' inputs and outputs) can help users form mental models for how to effectively use an application.

\textit{Example}: DALL-E provides curated examples of generated outputs and the prompts used to generate them. Adobe Photoshop provides pop-ups and tooltips to introduce the user to its Generative Fill feature.

\subsubsection{Understand the user's mental model*}
Conduct evaluations, such as interviews, to determine whether a user has formed a useful mental model of a generative AI application. One prompt that can be useful to ask is how they think the application provides a certain capability, which forces the user to articulate their theory of how the system works. The goal is not for the user to possess an accurate model, but rather, one that is useful for working effectively with the system. Furthermore, understanding the user’s mental model can also help you leverage their existing knowledge of similar applications to inform your design decisions.

\textit{Example}: In evaluating a Q\&A application, you might ask the user, ``how did the system answer your question about who the current President is?'' Answers such as, ``it looked it up on the web'' might indicate a need to educate users about hallucination issues. Users' existing mental models of other applications can also be useful to understand. For example, Github Copilot builds on users' mental models by following the same interaction pattern as its existing code completion features, which are familiar to many developers, hence easing their learning curve.

\subsubsection{Teach the AI system about the user}
LLMs are adept at tailoring their language to a target audience. Designers can induce these models to produce personalized responses to users – in essence, teaching the model about the user – by including additional prompt text such as, ``explain like I’m five'' or ``please give me a detailed, technical answer.'' Capturing the user’s expectations, behaviors, and preferences can improve the AI’s interactions with them. Users can also provide information about their background in UI outside of their prompt, which the application can then opaquely incorporate back into the prompt.

\textit{Example}: ChatGPT provides a form for ``Custom Instructions'' in which users provide answers to questions such as, ``Where are you based?'', ``What do you do for work?'', and ``What subjects can you talk about for hours?'' In this way, users teach ChatGPT about themselves in order to receive more personalized responses.

\subsection{Design for appropriate trust \& reliance}
Trustworthy generative AI applications are those that produce high-quality, useful, and (where applicable) factual outputs that are faithful to a source of truth. Calibrating users’ trust is crucial for establishing appropriate reliance: teaching users to scrutinize a model’s outputs for quality issues, inaccuracies, biases, underrepresentation, and other issues to determine whether they are acceptable (e.g. because they achieve a certain level of quality or veracity) or if they should be modified or rejected.

\subsubsection{Calibrate trust using explanations}
Be clear and upfront about what the application can and cannot do by explaining its capabilities and limitations. Teach users to be skeptical of potentially imperfect model outputs and help them understand when they can trust the system.

\textit{Example}: ChatGPT explains its capabilities (e.g. ``answer questions, help you learn, write code, brainstorm together'') and limitations (e.g. ``ChatGPT may give you inaccurate information. It's not intended to give advice.'') directly on its introduction screen. Google Bard provides a notice below the prompt field that states, ``Bard may display inaccurate or offensive information that doesn't represent Google's views.''

\subsubsection{Provide rationales for outputs}
Show the user why a particular output was generated by showing the model's ``chain of thought''~\cite{wei2022chain} or identifying the source materials used to generate it. For example, identifying source documents for answers expected to be factually correct or revealing the image sets that a text-to-image model was trained on can help users calibrate their trust.

\textit{Example}: Google Bard provides a list of sources it used to produce answers to questions. Adobe discloses that its Generative Fill feature was trained on ``stock imagery, openly licensed work, and public domain content where the copyright has expired''~\cite{adobe2023firefly}.

\subsubsection{Use friction to avoid overreliance}
Encourage the user to review and think critically about the generative model's outputs by introducing mechanisms that slow them down at key decision-making points. These mechanisms are known as cognitive forcing functions~\cite{buccinca2021trust}. Examples include offering multiple AI-generated options for the user to select from, highlighting uncertainty, or requiring the user to create content or render a decision before showing the AI’s output.

\textit{Example}: Google Bard displays multiple drafts for the user to review, which can encourage them to slow down and consider which drafts may be of lower or higher quality.

\subsubsection{Signify the role of the AI}
AI systems may act in different roles, such as ``tools,'' ``partners,'' ``analytics,'' or ``coaches''~\cite{mccomb2023focus}. These roles shape users' expectations of how the AI system fits into their workflow, such as the extent to which it takes initiative (e.g. by acting proactively vs. reactively) and agency (e.g. by directly manipulating an artifact vs. making suggestions or recommendations). The role that is signified shapes users' perceptions of the system: research by \citet{kim2023one} shows that AI ``tools'' are viewed as less genuine and caring than AI ``mediators'' or AI ``assistants.'' Anthropomorphic signifiers such as giving a human name to an AI, representing the AI with a human-like avatar, having the AI refer to itself using first-person pronouns, or showing an animated typing bubble to hide inference latency may all give users the false impression that they are interacting with a human. Although there is no clear consensus on the extent to which AI-infused user experiences should favor or discourage anthropomorphism~\cite{shneiderman2023on}, as a designer, it is crucial to clearly signify to the user when they are interacting with an AI system and what content was AI-generated.

\textit{Example}: Github Copilot's tagline is ``Your AI pair programmer'', which elicits the role of a partner. Copilot fulfills this role by proactively making suggestions as the user writes code. It also possesses a limited form of agency by making autocompletion suggestions directly in the user's code editor, although it requires the user to explicitly accept or reject those suggestions (e.g. by pressing tab or escape).

\subsection{Design for generative variability}
One distinguishing characteristic of generative AI systems is that they can produce multiple outputs that vary in character or quality, even when the user’s input does not change. This characteristic raises important design considerations: to what extent should multiple outputs be visible to users, and how might we help users organize and select amongst varied outputs?

\subsubsection{Leverage multiple outputs}
Take advantage of multiple outputs to help users produce the one that fits their needs. Multiple outputs can be exposed to the user or remain under the hood and instead allow the model to select the best option(s) to surface.

\textit{Example}: DreamStudio, DALL-E, and Midjourney all generate multiple distinct outputs for a given prompt; for example, DreamStudio produces four images by default and can be configured to produce up to 10. ChatGPT allows the user to regenerate a response to see more options.

\subsubsection{Visualize the user's journey}
Users may not be able to reproduce prior outputs. Capturing and displaying the user's history of outputs, along with the input parameters used (e.g. prompts, controls) can help them track their work. Also consider ways to guide the user to new output possibilities that they have not explored.

\textit{Example}: DreamStudio, DALL-E, and Midjourney all show a history of the user's inputs and resulting image outputs. Research prototypes extend the idea of ``visualizing the user's journey'' even further using visualization techniques to show unexplored parts of an output space~\cite{kreminski2022evaluating} (using dots overlaid on 2D histograms) or of a parameter configuration space~\cite{rost2023stable} (by showing configurations a user has and has not yet tried in a grid, with untried configurations rendered as placeholders).

\subsubsection{Enable curation \& annotation}
When a generative model produces multiple outputs, users may need to curate or annotate them. Curation may include collecting, filtering, or organizing outputs (possibly from the generation history). Annotation may include the ability to tag artifacts (e.g. ``I like this picture'') or make notes within an artifact (e.g. ``this line of code looks suspicious'').

\textit{Example}: DALL-E allows the user to mark images as favorites and store them within groups called collections. Users may create and name multiple public or private collections to organize their work.

\subsubsection{Draw attention to differences or variations across outputs}
A generative model can sometimes produce a set of similar outputs that are difficult to tell apart. When outputs are similar, tools that aid users in identifying the similarities and differences between multiple outputs can be useful.

\textit{Example}: DreamStudio, DALL-E, and Midjourney all display multiple outputs in a grid-like fashion to allow the user to identify differences, but fine-grained differences between outputs are not explicitly highlighted. A prototype source code translation interface by \citet{weisz2021perfection} visualizes the differences across multiple generated code translations through granular highlights, as well as interactively through a list of alternate translations.

\subsection{Design for co-creation}
Generative AI offers new co-creative capabilities. Help the user create outputs that meet their needs by providing controls that enable them to influence the generative process and work collaboratively with the AI.

\subsubsection{Help the user craft effective outcome specifications}
Generative AI has introduced a new interaction paradigm, intent-based outcome specification, in which users specify what they want but not how it should be produced. Orient the user to this new paradigm and assist them in prompting effectively to produce outputs that fit their needs.

\textit{Example}: Google Bard sends out a newsletter that includes a section called ``Prompt Engineering 101,'' which features tips and examples to help users improve their prompt writing. The IBM watsonx.ai Prompt Lab documentation includes a set of tips and examples to help the user understand how to improve their prompts.

\subsubsection{Provide generic input parameters}
Let the user control generic aspects of the generative process such as the number of outputs and the random seed used to produce those outputs. Generic controls apply across most use cases, independent of which model is used.

\textit{Example}: DreamStudio provides a slider for users to indicate the number of images they want to produce for a given prompt, along with an input field for random seed.

\subsubsection{Provide controls relevant to the use case and technology}
Let the user control parameters specific to their use case, domain, or model architecture. Some model architectures provide specialized means of control, such as semantic sliders for latent space models~\cite{liu2021neurosymbolic} or decoding strategy and temperature for LLMs. Other models have controls specific to the domain of the application (e.g. code, music, art).

\textit{Example}: AIVA allows the user to customize domain-specific characteristics of the musical compositions it generates, such as the type of ensemble and emotion.

\subsubsection{Support co-editing of generated outputs}
Allow both the user and the AI system to improve generated outputs. Although the user should retain agency and decision-making authority, the AI can co-create with the user to improve outputs, such as by providing suggestions, organizing outputs, or editing an artifact.

\textit{Example}: Adobe Photoshop exposes generative AI capabilities within the same design surface as its other image editing tools, enabling both the user and the generative AI model to co-edit an image.

\subsection{Design for imperfection}
Users must understand that generative model outputs may be imperfect according to objective metrics (e.g. untruthful or misleading answers, violations of prompt specifications) or subjective metrics (e.g. the user doesn't like the output). Provide transparency by identifying or highlighting possible imperfections, and help the user understand and work with outputs that may not align with their expectations.

\subsubsection{Make uncertainty visible}
Caution the user that outputs may not align with their expectations and identify detectable uncertainties or flaws. Provide disclaimers about the potential for imperfection, show the model’s confidence level when possible, and utilize ``I don’t know'' responses when confidence is low.

\textit{Example}: Google Bard's interface states, ``Bard may display inaccurate info, including about people, so double-check its responses.'' This disclaimer alerts the user to the possibility of uncertainties or imperfections in its outputs. A prototype source code translation interface proposed by \citet{weisz2021perfection} makes the generative model's uncertainty visible to the user by highlighting source code tokens based on the degree to which the underlying model is confident that they were correctly translated. These highlights help guide the user's attention toward places that require their review.

\subsubsection{Evaluate outputs using domain-specific metrics}
In some cases, the quality of a generative model's outputs can be assessed with measurable criteria. For example, answers to customer queries can be evaluated for faithfulness to source documents, and design mock-ups can be evaluated for adherence to a UI style guide.

\textit{Example}: Molecular candidates generated by CogMol~\cite{chenthamarakshan2020cogmol}, a prototype generative application for drug design, are evaluated with a molecular simulator to compute domain-specific attributes such as molecular weight, water solubility, and toxicity.

\subsubsection{Offer ways to improve outputs}
Provide ways for the user to fix imperfections and improve output quality, such as editing tools, an option to regenerate, or providing alternative outputs to select from.

\textit{Example}: DALL-E and DreamStudio allow users to refine outputs by erasing and regenerating parts of an image (inpainting) or generating new parts of the image beyond its boundaries (outpainting). Google Bard offers options for the user to modify outputs to be shorter, longer, simpler, more casual, or more professional.

\subsubsection{Provide feedback mechanisms}
Allow users to provide feedback on the quality of a model's output. Feedback may be implicit (e.g. the user makes edits to a generated artifact indicating potential issues) or it may be explicit (e.g. rating the quality with a thumbs up / thumbs down).

\textit{Example}: ChatGPT offers an option for the user to provide a thumbs up or thumbs down rating for its responses, along with open-ended textual feedback.

\end{document}